\documentclass[conference]{IEEEtran}
\IEEEoverridecommandlockouts

\usepackage[style=ieee,minnames=1,maxcitenames=1]{biblatex} 
\AtBeginBibliography{\small}

\usepackage{amsmath,amssymb,amsfonts}
\usepackage{algorithmic}
\usepackage{graphicx}
\usepackage{textcomp}
\usepackage{xcolor}

\IEEEaftertitletext{\vspace{-1\baselineskip}}

\mathchardef\mhyphen="2D 

\begin{document}

\renewcommand{\arraystretch}{1.125}

\title{Defectors: A Large, Diverse Python Dataset for Defect Prediction}
\vspace{-10em}
\author{
\IEEEauthorblockN{Parvez Mahbub}
\IEEEauthorblockA{
\textit{Dalhousie University}\\
parvezmrobin@dal.ca}
\and
\IEEEauthorblockN{Ohiduzzaman Shuvo}
\IEEEauthorblockA{
\textit{Dalhousie University}\\
ohiduzzamanshuvo@dal.ca}
\and
\IEEEauthorblockN{Mohammad Masudur Rahman}
\IEEEauthorblockA{
\textit{Dalhousie University}\\
masud.rahman@dal.ca}

}

\maketitle

\begin{abstract}

Defect prediction has been a popular research topic where machine learning (ML) and deep learning (DL) have found numerous applications.
However, these ML/DL-based defect prediction models are often limited by the quality and size of their datasets.
In this paper, we present Defectors, a large dataset for just-in-time and line-level defect prediction.
Defectors consists of $\approx$ 213K source code files ($\approx$ 93K defective and $\approx$ 120K defect-free) that span across 24 popular Python projects.
These projects come from 18 different domains, including machine learning, automation, and internet-of-things.
Such a scale and diversity make Defectors a suitable dataset for training ML/DL models, especially transformer models that require large and diverse datasets.
We also foresee several application areas of our dataset including defect prediction and defect explanation.
\end{abstract}


\section{Introduction}
A software defect is an incorrect step, process, or data definition in a computer program that prevents the program from working correctly~\cite{ieee-standard}.
Software defects are informally called software bugs.
They cost the global economy billions of dollars every year~\cite{Britton2013, zou2018practitioners}.
Despite the adoption of various software quality assurance (SQA) practices, defects may still sneak into official releases~\cite{thongtanunam2015investigating, thongtanunam2016revisiting}.
Furthermore, a recent work~\cite{pornprasit2022deeplinedp} shows that only $\approx$~3\% code lines of the whole release could lead to many of the bugs.
Therefore, prioritizing SQA efforts for highly risky areas of source code is essential to ensure the high quality of a software release.

Defect prediction has been a popular research topic for the last few decades. 
It identifies the defects in software code before releasing the software to end users.
It can also help prioritize the SQA efforts.
Defects can be predicted at different abstraction levels such as module~\cite{gong2021revisiting, yu2019empirical}, file~\cite{jiarpakdee2021practitioners, chen2020software}, method~\cite{shippey2019automatically}, and line~\cite{wattanakriengkrai2020predicting, jiang2013personalized, pornprasit2022deeplinedp, pornprasit2021jitline}.
In recent years, just-in-time (JIT) defect prediction~\cite{mcintosh2018fix, pascarella2019fine, huang2019revisiting, pornprasit2021jitline, hoang2019deepjit, hoang2020cc2vec, kamei2012large} also has gained significant attention, which predicts the defects just at the time of committing software changes.
Thus, a combination of line-level defect prediction and JIT defect prediction can provide a fine-grained location of a software defect.

Over the past few years, deep-learning models have been used for both line-level defect prediction~\cite{pornprasit2022deeplinedp} and JIT defect prediction~\cite{hoang2019deepjit, hoang2020cc2vec}.
Deep learning models provide state-of-the-art performance in various tasks of software engineering including bug localization~\cite{li2022fault, ni2022best} and bug explanation~\cite{mahbub2023bugsplainer}.
However, their performance in JIT defect prediction is sub-optimal.
Deep-learning-based tools such as DeepJIT~\cite{hoang2019deepjit} and CC2Vec~\cite{hoang2020cc2vec} cannot outperform simpler models such as logistic regression.
These models can be limited by the size and quality of their datasets.
First, the performance of ML/DL models often scales with the size of their dataset~\cite{zhu2016we, ccayir2021effect}.
However, most of the existing datasets used in defect prediction might not be large enough~\cite{keshavarz2022apachejit}.
Second, these datasets also suffer from the class imbalance problem containing only 5\%-26\% defective instances~\cite{mcintosh2018fix, keshavarz2022apachejit, kamei2012large}.
Such an imbalance could lead to sub-optimal performance with any deep-learning models.
Third, these datasets were constructed either from a small number of projects~\cite{mcintosh2018fix} or the projects from a single organization~\cite{keshavarz2022apachejit, fan2019impact}.
Such a choice limits the capability of these models to generalize their performances across different domains and organizations.

To mitigate the challenges with existing datasets, in this paper, we present \emph{Defectors} -- a large-scale dataset, containing both source code and their changes from 24 popular Python projects across 18 domains and 24 organizations.
We carefully identify defective source code files and their code changes, following five levels of noise filtration recommended in the literature.
Our dataset contains $\approx$~213K source code files ($\approx$~93K defective and $\approx$~120K defect-free).
It is suitable for training large models on the task of defect prediction that has the potential to provide high performance.

Defectors stands out from similar datasets in the following aspects.
\begin{enumerate}
    \item \textbf{Size:} To the best of our knowledge, Defectors is the largest defect prediction dataset and is twice in size as the previous largest dataset (i.e., $\approx$~106K)~\cite{keshavarz2022apachejit}.
    \item \textbf{Class Balance:} It maintains a near 1:1 ratio between defective and defect-free instances in the training set, where existing datasets contain only 5\%-26\% defective instances~\cite{mcintosh2018fix, keshavarz2022apachejit, kamei2012large}. 
    \item \textbf{Diversity in Application:} Defectors uses 24 projects from 18 application domains and 24 organizations, where existing datasets either use a small number of projects ($< 10$)~\cite{jiang2013personalized, mcintosh2018fix} or the projects from only one organization (e.g., Apache)~\cite{keshavarz2022apachejit, fan2019impact}.
    \item \textbf{Diversity in Platform:} Our dataset is based on Python projects, whereas nearly all existing datasets were constructed from Java-based projects. Thus, it diversifies the existing collection of defect prediction datasets.
\end{enumerate}

The dataset is publicly available at the following link: https://doi.org/10.5281/zenodo.7708984

\section{Dataset Description And Usage}
\label{sec: dataset-overview}

\begin{table}
    \centering
    \caption{Overview of our Defectors dataset}
    \label{tab: data-overview}
    \begin{tabular}{|l|l|p{4.5cm}|}
        \hline
        \textbf{Name} & \textbf{Type} & \textbf{Description} \\ \hline
        datetime & DateTime & Date and time with timezone from the committer \\ \hline
        commit & string & hash of the commit \\ \hline
        repo & string & Name of the project repository \\ \hline
        filepath & string & Path of a changed file in the \texttt{commit} \\ \hline
        content & string & For JIT defect prediction, the commit diff. For defect prediction in files, the content of the file after the \texttt{commit}. \\ \hline
        methods & List$<$string$>$ & Names of the methods in \texttt{filepath} which were changed in \texttt{commit}. \\ \hline
        induce\_bug &  List$<$int$>$ & A list of integers containing line numbers of the buggy code segment. \\ \hline
    \end{tabular}
    \vspace{-1em}
\end{table}

The Defectors dataset is one of the largest available datasets for defect prediction.
Table~\ref{tab: data-overview} shows an overview of the columns, their data types, and their descriptions.
Our dataset contains $\approx$ 213K source code files and their changes ($\approx$ 93K defective and $\approx$ 120K defect-free).
We store and distribute our dataset in Apache Parquet format, which is efficient, structured and compressed~\cite{Parquet}.

Our dataset is suitable for training large-scale models on software defect prediction.
The training dataset maintains a near 1:1 ratio for defective and defect-free data, whereas \emph{the test and validation datasets maintain the original distribution}.
The Defectors dataset can be adapted for method-level defect prediction as it contains names of the touched methods by a commit.
It can also be adapted for file-level defect prediction.
Since our dataset contains references to thousands of bug-fix commits, it can be extended for bug explanation~\cite{mahbub2023bugsplainer}.

\section{Dataset Construction}
\label{sec: dataset-contstruction}

\begin{table*}
    \centering
    \caption{Description of the projects used in Defectors}
    \label{tab: projects}
    \begin{tabular}{|l|l|l|r|r|r|}
        \hline
        \textbf{Project} & \textbf{Domain} & \textbf{Bug Report Management} & \textbf{Defective} & \textbf{Defect-free} & \textbf{Total}  \\ \hline \hline
        Lightning-AI/lightning  & deep-learning         & labeled PR        & 5,369 (48\%)  & 5,793 (52\%)  & 11,162 \\ \hline 
        ansible/ansible         & automation            & labeled PR        & 15,225 (35\%) & 28,340 (65\%) & 43,565 \\ \hline 
        apache/airflow          & automation            & labeled issue     & 4,788 (48\%)  & 5,166 (52\%)  & 9,954 \\ \hline 
        celery/celery           & task queue, messaging & labeled PR        & 805 (23\%)    & 2,674 (77\%)  & 3,479 \\ \hline 
        commaai/openpilot       & autonomouse driving   & labeled PR        & 695 (44\%)    & 885 (56\%)    & 1,580 \\ \hline 
        django/django           & web framework         & separate website  & 5,309 (47\%)  & 5,934 (53\%)  & 11,243 \\ \hline 
        encode/django-rest-framework & web framework    & labeled PR        & 269 (67\%)    & 134 (33\%)    & 403 \\ \hline 
        explosion/spaCy         & natural language processing & labeled PR  & 1,224 (50\%)  & 1,231 (50\%)  & 2,455 \\ \hline 
        getredash/redash        & data sciecne          & labeled PR        & 170 (24\%)    & 550 (76\%)    & 720 \\ \hline 
        getsentry/sentry        & logging               & PR title          & 13,791 (37\%) & 23,694 (63\%) & 37,485 \\ \hline 
        google/jax              & deep learning         & labeled issue     & 443 (61\%)    & 283 (39\%)    & 726 \\ \hline 
        home-assistant/core     & internet of things    & labeled PR        & 21,535 (60\%) & 14,413 (40\%) & 35,948 \\ \hline 
        huggingface/transformers & deep learning        & labeled issue     & 658 (36\%)    & 1,187 (64\%)  & 1,845 \\ \hline 
        localstack/localstack   & cloud, serverless     & labeled issue     & 925 (51\%)    & 879 (49\%)    & 1,804 \\ \hline 
        numpy/numpy             & data science          & labeled PR        & 466 (47\%)    & 527 (53\%)    & 993 \\ \hline 
        pandas-dev/pandas       & data science          & labeled PR        & 7,816 (50\%)  & 7,766 (50\%)  & 15,582 \\ \hline 
        psf/black               & development tool      & labeled issue     & 228 (41\%)    & 328 (59\%)    & 556 \\ \hline 
        pypa/pipenv             & development tool      & labeled issue     & 60 (50\%)     & 61 (50\%)     & 121 \\ \hline 
        Python/cPython          & programmin language   & labeled PR        & 1,769 (24\%)  & 5,507 (76\%)  & 7,276 \\ \hline 
        Python-poetry/poetry    & development tool      & labeled issue     & 1,283 (57\%)  & 949 (43\%)    & 2,232 \\ \hline 
        ray-project/ray         & machine learning, 
                                  deep learning         & labeled issue     & 7,405 (42\%)  & 10,336 (58\%) & 17,741 \\ \hline 
        scikit-learn/scikit-learn & machine learning    & labeled issue     & 2,565 (51\%)  & 2,434 (49\%)  & 4,999 \\ \hline 
        scrapy/scrapy           & crawling, scraping    & labeled issue     & 221 (52\%)    & 204 (48\%)    & 425 \\ \hline 
        ultralytics/yolov5      & deep learning, 
                                  image processing      & labeled issue     & 649 (58\%)    & 476 (42\%)    & 1,125 \\ \hline \hline
        \textbf{Total} & ‌\textbf{18 distinct domains} & & \textbf{93,668 (44\%)} & \textbf{119,751 (56\%)} & \textbf{213,419} \\ \hline
    \end{tabular}
    \footnotesize{
    \newline \newline
        $^*$labeled PR = bug fixing PRs are uniquely labeled, labeled issue = bug reporting issues are uniquely labeled, \\ separate website = bug reports are managed in a separate website, PR title = bug fixing PR titles have unique patter \\
        $^{**}$The values in the Defective, Defect-free, and Total columns are prior to creating train, validation, and test splits
    }
    \vspace{-2em}
\end{table*}

The Defectors dataset has been curated from 24 projects that span 18 domains and 24 organizations.
We capture past bug-inducing and bug-fixing changes from these projects to construct our dataset.

First, we select the projects based on their popularity and other quality aspects (e.g., consistent issue labelling).
Then, we identify the bug-fixing changes from these projects.
From them, we identify the bug-inducing changes using the SZZ algorithm~\cite{sliwerski2005szz}.
Then, we apply five levels of noise filtration to minimize the number of false positives identified by the SZZ algorithm.
Then, we sample defect-free code changes with 95\% confidence level and 5\% margin of error (see Section~\ref{subsec: defect-free}).
Finally, we store our dataset as train, validation, and test splits.
The following sections discuss the major steps of our dataset construction process.

\subsection{Project Selection}
\label{subsec: project-selection}

Most of the existing datasets used in defect prediction are constructed using Java projects.
To diversify the existing collection of datasets, we choose Python-based projects for our dataset.
Similar to existing studies~\cite{keshavarz2022apachejit}, we sort all the Python projects on GitHub in \emph{descending order} using their star counts.
Then, we manually investigate this ordered list of repositories sequentially to find mature and high-quality projects.
In each repository, we look for a consistent way of identifying the bug-fix pull requests (PR).
The following steps summarize our repository selection process\footnote{Accessed: December 2, 2022}.
\begin{enumerate}
    \item If a repository has less than 2000 PRs, we discard the project considering that it is not mature enough.
    \item For a mature repository, we identify all the bug-related labels (e.g., bug, bugfix) from the repository. 
    \item We find the PRs using these labels. If the number of such PRs is more than 100, then we accept the repository. We find 11 projects with consistently labelled PRs.
    \item If the number of bug-fix PRs is not sufficient, we find the issues that are associated with one of these labels and have a linked pull request resolving the issue. If the number of such issues is more than 100, then we accept the repository. We find 12 projects with consistently labelled and linked issues.
    \item If the number of neither PRs nor issues are sufficient, we read the titles of the first 200 PRs and look for any consistent patterns. We find that the titles of bug-fix PRs from one project start with a specific keyword (i.e., fix). The titles of bug-fix PRs from another project \emph{consistently} contain issue IDs from a different bug report management website. We accept these two projects for having such consistent patterns of labelling bug-fix PRs.
\end{enumerate}

By following the steps above, we select a total of 25 projects from various organizations and domains.
To find the first 25 projects matching these criteria, we had to investigate the first 100 repositories from the ordered list (based on descending star count).
Both the first and the second author individually selected these projects to avoid any mistakes.
Later, in our quality filtration stage, we discard a repository for not having enough bug-fix commits (see Section~\ref{subsec: filter}).
Table~\ref{tab: projects} enlists 24 projects along with their domain, bug report management system, and the ratio of defective and defect-free code changes in our dataset.

\subsection{Bug Fixing Commit Collection}
During project selection, we check their bug report management system.
For projects where bug-fix PRs are directly labelled with appropriate labels, we collect the labelled PRs.
For projects where issue reports are labelled with appropriate labels, we first collect the issues and then collect their associated PRs.
Using these two approaches, we collect $\approx$~39K bug-fix PRs from 23 projects.
For the remaining two projects, we use slightly different approaches. \texttt{getsentry/sentry} follows a pattern where all bug-fix PRs start with the keyword -- \textit{fix}.
Therefore, we collect the PRs with such a pattern.
Finally, \texttt{django/django} uses a separate website\footnote{https://code.djangoproject.com} to manage their issues.
In GitHub, its PRs contain corresponding issue IDs from that site.
We first collect the closed bug reports from the issue site and then capture the corresponding PRs from GitHub.
Once we have the bug-fix PRs, we collect their merge commits as the bug-fix commits.
This way, we collect $\approx$ 51K bug-fix commits from 25 projects.

\subsection{Bug Inducing Commit Collection}
In this step, we capture the bug-inducing commits from the bug-fix commits using the SZZ algorithm~\cite{sliwerski2005szz}.
Most of the studies in defect prediction~\cite{mcintosh2018fix, da2016framework,jiang2013personalized, keshavarz2022apachejit, fan2019impact} use the SZZ algorithm to identify the bug-inducing changes.
We use an implementation of the SZZ algorithm by the PyDriller tool~\cite{spadini2018pydriller}.
This implementation takes a commit as the input and returns a list of commits that last changed the deleted lines in the input commit.

\subsection{Bug Inducing Commits Filtration}
\label{subsec: filter}
The SZZ algorithm yields a considerable amount of false positives, i.e., identifies defect-free commits as defective commits~\cite{fan2019impact}.
Thus, we apply a series of filtration inspired by the literature to minimize the number of false positives.

\subsubsection{Filtration Using The Number of Linked Bug Inducing Commits} SZZ often links several bug-inducing commits to a single bug-fix commit.
This suggests that a bug could occur due to non-coherent changes in hundreds or even thousands of files, which is impractical.
Therefore, existing studies discard the bug-fix commits that are linked to too many bug-inducing commits~\cite{kamei2012large, mcintosh2018fix, keshavarz2022apachejit}.
Let $inducer\mhyphen count$ be the number of bug-inducing commits linked to a single bug-fix commit.
\textcite{keshavarz2022apachejit} suggest using Equation~\ref{eq: thresh} as a threshold.
\begin{equation}
\label{eq: thresh}
    thresh(X) = mean(X) + std(X)
\end{equation}
In this work, we use a threshold of 14, derived from the above equation to filter out the noisy bug-fix commits.

\subsubsection{Filtration Using The Number of Linked Bug-fix Commits}
Let $fixer\mhyphen count$ be the number of bug-fix commits linked to a single bug-inducing commit.
If a bug-inducing commit has a $fixer\mhyphen count$ greater than one, it suggests that the commit induced multiple bugs in the project, which are reported by multiple bug reports and are fixed by multiple bug-fix commits.
Similar to $inducer\mhyphen count$, we apply Equation~\ref{eq: thresh} to $fixer\mhyphen count$ as well.
We thus discard the bug-inducing commits having a $fixer\mhyphen count$ higher than 7, which is derived from the above equation.

\subsubsection{Filtration Using The Size of Changed Code}
If a commit changes a large number of lines or files, it indicates that the commit might contain tangled changes.
During our manual analysis, we found some commits that modified even up to 1,000 files.
Often these commits indicate some administrative tasks, for instance, merging several related projects into a single repository.
Therefore, existing studies~\cite{mcintosh2018fix, keshavarz2022apachejit} filter out the commits that are large.
Similarly, we filter out the bug-inducing commits that have more than 1000 changed lines or have touched more than 100 files.

\subsubsection{Filtration Using File Type}
In this paper, we focus specifically on Python source code.
Such a language constraint makes it easy to perform static analysis on source code.
It also helps us capture the structural information from source code (e.g., changed methods).
Even though Python is the main language of all of our projects, they contain a small fraction of non-Python files (e.g., configuration files).
Thus, we filter out the commits that do not modify any Python file.

\subsubsection{Filtration Using The Nature of Change}
All the changes in a source code file might not be bug-inducing. 
For instance, changes in the comments or code formatting generally do not introduce new bugs.
As done by existing studies~\cite{keshavarz2022apachejit, da2016framework, kim2006automatic}, we thus discard such trivial changes.
Trivial changes do not modify the abstract syntax tree (AST) of the source code.
Therefore, to identify trivial changes, we compare the AST of the source code before the commit to the AST after the commit.
If both ASTs are the same, then the commit performs a trivial change, which is discarded.
Our implementation of this filtration is tolerant of syntax errors.
That is, if the source code is not syntactically correct, our implementation will still generate partial AST for comparison.

After completing all these filtrations, we find that one project, namely -- \texttt{freqtrade/freqtrade}, contains only one bug-fix commit.
We thus discard the project and keep the remaining 24 repositories in our dataset.

\subsection{Collecting and Sampling Defect-free Commits}
\label{subsec: defect-free}

We collect all the commits within the date range of the defective (i.e., bug-inducing) commits from the same project.
Then, we separate the defective commits from the defect-free commits using the commit hashes.
In software projects, defect-free commits often outnumber defective commits by a large margin.
Therefore, we down-sample the defect-free commits to ensure a near 1:1 class ratio.
For each project, we sample defect-free commits with a 95\% confidence level and a 5\% margin of error.
If this sample size is less than the number of defective commits, then we increase the sample size to achieve parity.
Finally, we discard the defect-free commits that do not modify any Python file.

\subsection{Formalizing The Dataset}
We formalize our dataset by targeting two tasks --  just-in-time (JIT) defect prediction (i.e., defect prediction on commit diff) and defect prediction from source files.
For JIT defect prediction, the input is \emph{git diff}, and for defect prediction from source files, the input is the content of the file after the commit.
In both cases, the output is the list of added (i.e., defective) line numbers if the commit is defective.
Otherwise, the output is an empty list.
For both variants, we make train-validation sets based on both random and time-wise splitting approaches.
The training splits maintain a near 1:1 ratio of defective and defect-free instances, where \emph{test and validation splits maintain the original distribution}.
In particular, $\approx$~7\% files in the codebase are defective and $\approx$~4\% lines are defective in the defective files.
These variants of our dataset could help to effectively train and comprehensively evaluate large-scale defect prediction models.

\section{Limitations}
Similar to former studies, we use SZZ~\cite{sliwerski2005szz} algorithm to identify the bug-inducing commits.
Despite its limitations~\cite{da2016framework, fan2019impact}, SZZ is the most used algorithm to identify bug-inducing commits.
To minimize the effect of mislabelling by SZZ, we follow several filtration steps from the literature~\cite{da2016framework, mcintosh2018fix, keshavarz2022apachejit} that have been shown to be effective.
In this work, we limit our focus to a single language (i.e., Python).
However, our dataset construction approach is language agnostic (see Section~\ref{sec: dataset-contstruction}) and can be adapted to any programming language.

\section{Similar Datasets}

\textcite{kamei2012large} perform a large-scale study on change-level defect prediction using six open-source and five closed-source projects.
They use the original SZZ algorithm~\cite{sliwerski2005szz} to annotate the bug-inducing changes.
However, their dataset is not publicly available.
\textcite{mcintosh2018fix} conduct a time-series analysis on JIT defect prediction using two rapidly evolving projects.
\textcite{jiang2013personalized} attempt to personalize defect prediction for different developers.
Their dataset consists of only six open-source projects and is not publicly available.
\textcite{keshavarz2022apachejit} present ApacheJIT, a large dataset for JIT defect prediction.
This dataset is derived from 14 open-source Java projects and contains $\approx$~106K software revisions.
Despite their large size, this dataset contains projects from a single organization.
Such a choice might limit the generalizability of a model trained on their dataset.

All of these datasets use heuristic-based approaches that look for certain keywords (e.g., fix) or issue numbers in commit messages to identify bug-fixing changes.
However, \textcite{yatish2019mining} mention that such approaches produce a significant amount of mislabeled data.
They suggest using bug-fix labels directly provided by the developers.
In a case study of nine projects, they show that this approach leads to improvement in prediction accuracy.
In our work, we use projects where bug-fix labels are directly provided by the developers.
The distinction between our dataset and that of \textcite{yatish2019mining} is that our dataset is more diverse than theirs in terms of repositories (9 vs. 24 repositories) and organizations (1 vs. 24 organizations).
Furthermore, all of the aforementioned datasets are constructed from Java projects whereas our dataset is constructed from Python projects.

\section{Conclusion}
Among various defect prediction approaches, line-level defect prediction and just-in-time (JIT) defect prediction offer the most precise predictions.
The deep-learning models have strong potential to support such predictions.
However, the performance of these models could still be limited by the lack of a large, diverse, and balanced dataset.
In this paper, we present Defectors -- a large dataset suitable for both line-level defect prediction and JIT defect prediction.
Our dataset can be used to train large deep-learning models that are precise in their defect predictions and could be generalizable.

\section*{Acknowledgement}
This work was supported by Dalhousie University and Mitacs Accelerate International Program. We would like to thank Avinash Gopal, Ben Reaves, and Massimiliano Genta from our industry partner -- \emph{Metabob Inc}. 

\printbibliography

\end{document}